\documentclass[prb,twocolumn,superscriptaddress]{revtex4-2}
\usepackage{amsmath,amssymb,amsfonts,mathrsfs} 	
\usepackage{graphicx}
\usepackage{subfigure}
\usepackage{xcolor}
\usepackage[normalem]{ulem}

\renewcommand{\[}{\begin{equation}}
\renewcommand{\]}{\end{equation}}
\def\rr{}
\def\bea{\begin{eqnarray}}
\def\eea{\end{eqnarray}}
\def\nn{\nonumber\\}

\newcommand{\emi}[1]{{\rm e}^{-i #1}}
\newcommand{\ei}[1]{{\rm e}^{i #1}}

\newcommand{\EE}{{\cal E}}
\newcommand{\FF}{{\cal F}}

\renewcommand{\j}{{\bf j}}
\newcommand{\p}{{\bf p}}

\renewcommand{\k}{{\bf k}}

\newcommand{\kk}{\mbox{\boldmath$\kappa$}}
\newcommand{\vc}{V_{\rm cell}}
\renewcommand{\v}{{\bf v}}
\renewcommand{\r}{{\bf r}}

\def\EEE{\mbox{\boldmath${\cal E}$}}

\newcommand{\equ}[1]{Eq.~(\ref{#1})}
\newcommand{\eqs}[2]{Eqs.~(\ref{#1}) and (\ref{#2})}

\def\bra#1{\langle#1\vert}
\def\ket#1{\vert#1\rangle}
\def\ev#1{\langle#1\rangle}
\def\me#1#2#3{\langle#1| \, #2 \, |#3\rangle}
\def\runtime{(\the\time)\qquad\the\month/\the\day/\the\year}
\def\today
 {\count10=\year\advance\count10 by -2000 \number\day--\ifcase
  \month \or Jan\or Feb\or Mar\or Apr\or May\or Jun\or
             Jul\or Aug\or Sep\or Oct\or Nov\or Dec\fi--\number\count10}

\def\hour{\count10=\time\count11=\count10
\divide\count10 by 60 \count12=\count10
\multiply\count12 by 60 \advance\count11 by -\count12\count12=0
\number\count10 :\ifnum\count11 < 10 \number\count12\fi\number\count11}

\begin{document}

\title{Nonadiabatic quantum geometry and optical conductivity}

\author{Raffaele Resta}
\email{resta@iom.cnr.it.it}
\affiliation{CNR-IOM Istituto Officina dei Materiali, Strada Costiera 11, 34149 Trieste, Italy}
\affiliation{Donostia International Physics Center, 20018 San Sebasti{\'a}n, Spain}


\begin{abstract} The ground-state quantum geometry is at the root of several static and adiabatic properties, while genuinely dynamic properties are routinely addressed via Kubo formul\ae, whose essential entries are the excited states. It is shown here that the ground-state metric-curvature tensor evolves in time by means of a causal unitary operator, which by construction elucidates the geometrical effect of the excited states in compact form. In the condensed-matter case the generalized tensor encompasses the whole conductivity tensor at arbitrary frequencies in both insulators and metals, with the exception of the Drude term in the metallic case; the latter is shown to be eminently nongeometrical.
\end{abstract}



\maketitle 

\section{Introduction}

Quantum geometry made its appearance in electronic-structure theory the early 1980s, in the form of an entity nowadays called Berry curvature \cite{Berry84,Niu84}. Since then, the quantum geometry of the ground state  constitutes a conceptual foundation and a computational  framework for several---apparently disparate---static and adiabatic  properties in condensed-matter \cite{Vanderbilt,Xiao10,Nagaosa10,rap_a37}  and in molecules \cite{rap168}.  It has been challenging, instead, to relate quantum geometry to genuinely nonadiabatic properties, which in a Kubo formulation involve excited states as well. 

Some pioneering results in this direction appeared recently, all of them  in a band-structure framework addressing the mean-field crystalline case \cite{Ahn22,Nagaosa22,Komissarov24,Verma25}. The geometrical nature of a dynamical response property is  perspicuous in the time domain, where the metric-curvature tensor is made to evolve
 by means of a unitary causal operator, \equ{o1} below, which accounts for the effect of the excited states in a meaningful form. The geometrical tensor so defined allows retrieving the full conductivity tensor---real and imaginary parts, longitudinal and transverse terms---in both insulators and metals, with one exception in the metallic case: the Drude term. The latter is shown to be in fact the {\it nongeometrical} term in the adiabatic response to a dc field. 

{\rr Most of the  electronic-structure literature addressing geometrical issues is formulated for noninteracting electrons in a crystalline system \cite{Vanderbilt,Nagaosa10,rap_a37,Ahn22,Nagaosa22,Komissarov24,Verma25}. Therein the state vectors are the cell-periodic Bloch orbitals, having a parameter dependence on the Bloch vector $\k$; intensive observables are expressed as Brillouin-zone (BZ) integrals in the insulating case, and as Fermi-volume integrals in the metallic case. Here instead I adopt the  many-body formulation of quantum geometry \cite{Niu84,Ortiz94,Xiao10,rap165,rap169}. Although it  addresses  in principle even systems with disorder and correlation,  its main virtue here is that it allows  for compact and very transparent notations; all geometrical  formula\ae\ can be easily converted when needed in their more prolix Bloch counterparts following a set of simple rules \cite{rap165}. This is shown in detail for the present nonadiabatic case in Sec. IV.
}

\section{Adiabatic quantum geometry}

{\rr The role of the familiar $\k$ parameter in Bloch geometry \cite{Vanderbilt,rap_a37}  in the many-body case is taken instead by the parameter $\kk$, called ``flux'' or ``twist'',  defined next; both $\k$ and $\kk$ have the dimensions of an inverse length.}
One considers a system of $N$ interacting $d$-dimensional electrons in a cubic box of volume $L^d$, and the family of many-body Hamiltonians parametrized by $\kk$: \[ \hat{H}_{\kk} = \frac{1}{2m} \sum_{i=1}^N \left(\p_i + \frac{e}{c}{\bf A}(\r_i) + \hbar \kk \right)^2 + \hat{V}, \label{kohn} \] where $\hat{V}$ includes the one-body potential (possibly disordered) and electron-electron interaction; the vector potential summarizes all intrinsic time-reversal  breaking terms, as e.g. those due to a coupling to a background of local moments. 
We assume the system to be macroscopically homogeneous; the eigenstates are normalized to one in the hypercube of volume $L^{Nd}$. 
The thermodynamic limit $N \rightarrow \infty$, $L \rightarrow \infty$, $N/L^d$ constant is understood throughout this work. 

We impose  Born-von-K\`arm\`an  periodic boundary conditions: the many-body wavefunctions are periodic with period $L$ over each electron coordinate $\r_i$ independently; the
potential $\hat{V}$ and the vector potential ${\bf A}(\r)$ enjoy the same periodicity. 
 If the eigenvalue problem is instead solved for the same  Hamiltonian by adopting  standard ${\mathscr L}^2$ {\rr (i.e. square-integrable)} boundary conditions, one may  address some geometrical properties other than conductivity in molecules \cite{rap168}. The flux $\kk$ corresponds to perturbing the Hamiltonian with a vector potential  $\hbar c \kk /e$, constant in space; if it is constant even in time, then its occurrence is a pure gauge.

The symbols  $\hat{H}$, $\ket{\Psi_n}$ and   $E_n$ will refer to $\kk=0$ quantities;
the velocity operator at $\kk=0$ is \[ \hat{\v} = \frac{1}{m} \sum_{i=1}^N \left[\p_i + \frac{e}{c}{\bf A}(\r_i)  \right] = \frac{1}{\hbar} \partial_{\kk} \hat{H} . \] 
The Berry curvature at $\kk=0$ is 
\[ \Omega_{\alpha\beta} =- 2  \, \mbox{Im } \ev{\partial_{\kappa_\alpha} \Psi_0 |\partial_{\kappa_\beta} \Psi_0}  , \label{berry} \] where Greek subscripts indicate Cartesian coordinates; sum over repeated subscripts is implicitly understood in the following.
The intrinsic anomalous Hall conductivity is \[ \sigma_{\alpha\beta}^{(-)}(0) = - \frac{e^2}{\hbar  L^d}  \Omega_{\alpha\beta} ; \label{ahe} \] the expression holds for metals and insulators, in either 2$d$ or 3$d$. The equivalence of \eqs{berry}{ahe} to the corresponding Kubo formula is shown in \eqs{s3}{cc}. 

For $d=2$ and in the insulating case \equ{ahe} is topologically quantized, although in the $L \rightarrow \infty$ limit only. The single-$\kk$ expression $2\pi \Omega_{xy}/L^2$ converges in fact  to the Chern number \cite{rap135}, while instead the $\kk$-averaged value is quantized at any $L$ \cite{Niu84,Xiao10,rap165}.

\section{Nonadiabatic quantum geometry}

The Berry curvature is naturally endowed with a symmetric part and with a time dependence by defining \bea F_{\alpha\beta}(t) &=& \ev{\partial_{\kappa_\alpha} \Psi_0 | \tilde{U}(t) |\partial_{\kappa_\beta} \Psi_0}, \nn \tilde{U}(t)  &=& \theta(t) \sum_n \ket{\Psi_n}\emi{\omega_{0n} t}\bra{\Psi_n} .\label{o1} \eea where $\omega_{0n} = (E_n - E_0)/\hbar$; this expression is not gauge-invariant. Therefore it is mandatory to evaluate the $\kk$-derivatives  in the parallel-transport gauge \cite{Vanderbilt}: {\rr \[ \ket{\partial_{\kk} \Psi_0}  = -\sum_{n\neq 0} \ket{\Psi_n} \frac{\me{\Psi_n}{\hat{\v}}{\Psi_0}}{ \omega_{0n}} , \]
i.e. setting $\ev{\Psi_0 | \partial_{\kk} \Psi_0} = 0 $. One thus gets}
\[ F_{\alpha\beta}(t) = \theta(t) \sum_{n\neq 0}    \frac{\langle \Psi_0 | \hat v_\alpha | \Psi_n \rangle \langle
\Psi_n | \hat v_\beta | \Psi_0 \rangle}{\omega_{0n}^2} \emi{\omega_{0n} t} . \label{F}  \] By  construction, the ground-state Berry curvature is retrieved as  \[ \Omega_{\alpha\beta} = - 2 \, \mbox{Im } F_{\alpha\beta}(0^+) , \label{b2} \] while $\mbox{Re } F_{\alpha\beta}(0^+) $ is the quantum metric evaluated in the parallel-transport gauge.

{\rr Owing to the Souza-Wilkens-Martin sum rule the ground-state metric is divergent in metals \cite{Souza00,rap_a33}, 
and therefore for the time being only the insulating case is addressed. It will be shown below how to deal with the metallic case as well.} In the present approach, both the metric and the curvature evolve causally by means of  the unitary operator $\tilde{U}(t)$.

The Fourier transform of \equ{F} is \[ \FF_{\alpha\beta}(\omega) = \sum_{n\neq 0} \frac{\langle \Psi_0 | \hat v_\alpha | \Psi_n \rangle \langle \Psi_n | \hat v_\beta | \Psi_0 \rangle}{\omega_{0n}^2}  \frac{i}{\omega- \omega_{0n} + i \eta} , \label{t1}\] where $\eta \rightarrow 0^+$. In order to compare $\FF_{\alpha\beta}(\omega)$ to the Kubo formul\ae\ for optical conductivity it is expedient to adopt the notation \[ \langle \Psi_0 | \hat v_\alpha | \Psi_n \rangle \langle
\Psi_n | \hat v_\beta | \Psi_0 \rangle = {\cal R}_{n,\alpha\beta} + i \, {\cal I}_{n,\alpha\beta} , \] where the real term ${\cal R}_{n,\alpha\beta}$ is symmetric and the imaginary one ${\cal I}_{n,\alpha\beta} $ is antisymmetric.

The symmetric and antisymmetric parts of the Fourier transform of  $F_{\alpha\beta}(t)$ are:
\bea {\cal F}^{(+)}_{\alpha\beta}(\omega) &=&  \sum_{n \neq 0}  \frac{{\cal R}_{n,\alpha\beta} }{\omega_{0n}^2} \left[\pi  \delta(\omega - \omega_{0n}) + \frac{i}{\omega- \omega_{0n}} \right],  \label{l2} \\ {\cal F}^{(-)}_{\alpha\beta}(\omega) &=&  \sum_{n \neq 0}  \frac{{\cal I}_{n,\alpha\beta} }{\omega_{0n}^2} \left[i \pi  \delta(\omega - \omega_{0n}) -  \frac{1}{\omega- \omega_{0n}} \right] , \eea 
Two of the above expressions are immediately  related to  standard Kubo formul\ae\ for $\omega > 0$, \eqs{s1}{s4}:
\bea  \mbox{Re } \sigma_{\alpha\beta}^{(+)}(\omega) &=& \frac{e^2}{\hbar  L^d}  \omega \, \mbox{Re }{\cal F}^{(+)}_{\alpha\beta}(\omega)  ; \label{longi} \\   \mbox{Im } \sigma_{\alpha\beta}^{(-)}(\omega)  &=&\frac{e^2}{\hbar  L^d}  \omega \, \mbox{Im }{\cal F}^{(-)}_{\alpha\beta}(\omega) . \label{main} \eea 
The independent-electron version of \eqs{l2}{longi} was previously arrived at by following a very different logical path \cite{Nagaosa22,Ahn22,Komissarov24,Verma25}.

The remaining Kubo formul\ae\ are also obtained in terms  of ${\cal F}_{\alpha\beta}(\omega)$ as $\omega$-integrals via the Kramers-Kronig relationships.
The real part of the transverse conductivity is thus
\[    \mbox{Re } \sigma_{\alpha\beta}^{(-)}(\omega) = \frac{e^2}{\hbar  L^d}  \frac{2}{\pi} \int_0^\infty d\omega' \; \frac{\omega'^2\mbox{Im } \FF_{\alpha\beta}^{(-}(\omega')}{\omega'^2 - \omega^2}  \label{k2} , \]  yielding in the dc case 
\[    \sigma_{\alpha\beta}^{(-)}(0) = \frac{2 e^2}{\pi \hbar  L^d}  \int_0^\infty d\omega \; \mbox{Im } \FF_{\alpha\beta}^{(-)}(\omega)  , \] consistently with \eqs{ahe}{b2}.

The imaginary part of the longitudinal conductivity is \[    \mbox{Im } \sigma_{\alpha\beta}^{(+)}(\omega) = - \frac{e^2}{\hbar  L^d}  \frac{2 \omega}{\pi} \int_0^\infty d\omega' \; \frac{\omega'\mbox{Re  } \FF_{\alpha\beta}^{(+)}(\omega')}{\omega'^2 - \omega^2}  \label{k1} ; \] in insulators this quantity is related to the static polarizability \cite{Komissarov24}  as  \bea \chi_{\alpha\beta}(0)  &=& \lim_{\omega \rightarrow 0} \frac{\mbox{Im }\sigma^{(+)}_{\alpha\beta}(\omega)}{\omega} \nn &=&  - \frac{2 e^2}{ \pi\hbar  L^d}   \int_0^\infty d\omega \; \frac{\mbox{Re  } \FF_{\alpha\beta}^{(+)}(\omega)}{\omega} . \eea Even this expression is proved to be equivalent to the corresponding Kubo formula,  \equ{ss}.

In the metallic case a comparison with the Kubo formula for longitudinal conductivity, \eqs{cond}{s1},  shows that \equ{longi}  yields the regular term only and lacks the Drude contribution. It has been observed above that $\mbox{Re } F_{\alpha\beta}(0^+) $---i.e. the quantum metric---diverges in metals: this divergence owes in fact to the Drude term in the Souza-Wilkens-Martin sum rule \cite{Souza00,rap_a33}. {\rr A regular metric can be {\it defined} even in the metallic case by means of  the said sum rule 
but {\it excluding}  the Drude term; such metric  coincides }indeed with $\mbox{Re } F_{\alpha\beta}(0^+) $.

The Drude term accounts for the longitudinal adiabatic response of the many-electron system to a dc field; it
can be expressed in the same $\kk$-space whose geometry we are investigating here, but it will be shown that in the present framework it is eminently {\it nongeometrical}.  

Quite generally, the adiabatic evolution of an observable is comprised of a nongeometrical and a geometrical term: in the present case of dc conductivity, the longitudinal one is the nongeometrical term and the transverse one is the geometrical term.

The adiabatic observable is the macroscopic current density, whose operator is \[ \hat{\j} = - \frac{e}{L^d} \hat{\v} =  - \frac{e}{\hbar L^d} \partial_{\kk} \hat{H}. \] If the Hamiltonian acquires a time dependence, the $\kk$-derivative of the energy is \bea \partial_{\kappa_\alpha} \me{\Psi_t}{\hat{H}_t}{\Psi_t} &=& \me{\Psi_t}{\partial_{\kappa_\alpha} \hat{H}_t}{\Psi_t}  \label{exa} \\ &+& i \hbar (\ev{\partial_{\kappa_\alpha}  \Psi_t | \dot\Psi_t} - \ev{\dot\Psi_t | \partial_{\kappa_\alpha}  \Psi_t} ) . \nonumber \eea The adiabatic theorem states that ``when the change of the Hamiltonian in time is made infinitely slow, the system, when started from a stationary state, passes through the corresponding stationary states for all times'' \cite{Kato50}. If we set $\ket{\Psi_t}$ = $\ket{\Psi_0}$  at $t=0$, the adiabatic limit  of \equ{exa} yields \[ \me{\Psi_0}{\partial_{\kappa_\alpha} \hat{H}}{\Psi_0} = \partial_{\kappa_\alpha} E_0 - \hbar \,\Omega_{\alpha\beta} \,\dot\kappa_\beta , \label{smart} \] where the Hellmann-Feynman theorem has also been used; the entries of \equ{smart} are the instantaneous ground state and energy. The two terms in the adiabatic evolution are clearly the nongeometrical and geometrical one, respectively.

It has been observed above that $\kk$ is actually a vector potential; therefore $\dot\kk = -e \EEE/\hbar$ when a macroscopic field $\EEE$ is switched on. By considering its geometrical term only, \equ{smart} yields \[ j_\alpha =  \frac{e}{L^d} \Omega_{\alpha\beta} \,\dot\kappa_\beta = - \frac{e^2}{\hbar  L^d}  \Omega_{\alpha\beta} \, \EE_\beta , \] which indeed coincides with \equ{ahe}. If we consider instead the nongeometrical term only in \equ{smart} the time-derivative of the current is \[ \partial_t  j_\alpha(t) = -\frac{e}{\hbar L^d} \frac{\partial^2 E_0}{\partial t \partial \kappa_\alpha}  =  \frac{e^2}{\hbar^2 L^d} \frac{\partial^2 E_0}{\partial \kappa_\alpha \partial \kappa_\beta} \EE_\beta  ,  \label{acce} \]  clearly showing that the many-electron system in a dc field undergoes free acceleration (in absence of extrinsic dissipation mechanisms). The causal Fourier transform of \equ{acce} leads to the famous Kohn's formula for longitudinal dc conductivity \cite{Kohn64}: 
 \[ \sigma^{(\rm Drude)}_{\alpha\beta}(\omega) = \frac{D_{\alpha\beta}}{\pi} \frac{i}{\omega + i\eta} , \quad  D_{\alpha\beta}  = \frac{\pi e^2}{\hbar^2 L^d} \frac{\partial^2 E_0}{\partial \kappa_\alpha \partial \kappa_\beta} .\label{omega1} \]

Therefore---as stated above---the transverse dc conductivity is geometrical and is in fact proportional to $ \mbox{Im } F_{\alpha\beta}(0^+)$; the longitudinal one is nongeometrical and out of reach of our tensor $F_{\alpha\beta}(t)$ as a matter of principle.

Despite being eminently a nongeometrical quantity, the Drude weight $ D_{\alpha\beta}$ can nonetheless be retrieved from our geometric tensor by means of the $f$-sum rule, \equ{fsum}: \[ D_{\alpha\beta} = D^{(\rm free)}_{\alpha\beta} -   \frac{2 e^2}{\hbar  L^d} \int_0^\infty d \omega \; \omega \mbox{Re } \FF_{\alpha\beta}^{(+)}(\omega) , \]  where \[D^{(\rm free)}_{\alpha\beta} =\frac{\pi e^2 N}{m L^d}\delta_{\alpha\beta}  \] is the Drude weight of an electron gas, i.e. an interacting many-electron system in a flat potential \cite{GiulianiVignale}.

\section{Conversion to Bloch formul\ae}
 
In the special case of a crystalline system of noninteracting electrons in a Kohn-Sham framework the ground-state $\ket{\Psi_0}$ is a Slater determinant of Bloch orbitals $\ket{\psi_{j\k}} = \ei{\k \cdot \r} \ket{u_{j\k}}$ with eigenvalues $\epsilon_{j\k}$, normalized in the crystal cell of volume $\vc$. The $\ket{u_{j\k}}$ orbitals are eigenstates of the one-body Hamiltonian  \[ H_\k = \frac{1}{2m} \left( {\bf p} +\frac{e}{c} {\bf A}(\r)  + \hbar \k  + \hbar \kk \right)^2 + V(\r) , \]  hence a $\kk$-derivative coincides with a $\k$-derivative. The $n$-th state  is a monoexcited determinant where the occupied orbital $\ket{u_{j\k}}$ is replaced by the unoccupied $\ket{u_{j'\k}}$ one; the vertical transition frequency is $\omega_{jj'\k} = (\epsilon_{j'\k} - \epsilon_{j\k})/\hbar$. The many-body matrix elements of $\hat{\v}$  are transformed---owing to the Slater-Condon rules---into the matrix element of the corresponding one-body velocity $\v = \partial_\k H_\k / \hbar$, evaluated at $\kk=0$. If $\epsilon_{\rm F}$ is the Fermi level, \equ{t1} becomes  \[ \FF_{\alpha\beta}(\omega) = \!\!\!\!\!\!\! 
\sum_{\substack{\epsilon_{j\k} < \epsilon_{\rm F} \\ \epsilon_{j'\k} > \epsilon_{\rm F }}} 
\!\!\!  \frac{\langle u_{j\k}  |  v_\alpha |  u_{j'\k}  \rangle  \langle u_{j'\k}  | v_\beta | u_{j\k} \rangle}{\omega^2_{jj'\k}}  \frac{i}{\omega- \omega_{jj'\k} + i \eta} \label{t2} ;\] the formula  is  given here per spin channel.

As usual when dealing with conductivities, the singular function $\FF_{\alpha\beta}(\omega)$ becomes a regular one in the thermodynamic limit $L \rightarrow \infty$. The limit can be explicitly performed in the crystalline case, where the sum over $\k$ converges to a BZ integral (Fermi-volume integral in the metallic case):
\bea \FF_{\alpha\beta}(\omega) &=&   \frac{\vc}{(2\pi)^d} \sum_{jj'} \int_{\rm BZ} \!\!\!\! d \k \; \frac{i \, f_{j\k} (1- f_{j'\k} )}{\omega- \omega_{jj'\k} + i \eta} \nn &\times& \frac{\langle u_{j\k}  |  v_\alpha |  u_{j'\k}  \rangle  \langle u_{j'\k}  | v_\beta | u_{j\k} \rangle}{\omega^2_{jj'\k}}   ,\label{t3}  \eea where $ f_{j\k}$ is the Fermi occupancy factor.

Analogously to  the many-body original definition, \equ{o1}, the geometrical content of the tensor is clearly apparent in the time domain:
\bea F_{\alpha\beta}(t) &=&\frac{\vc}{(2\pi)^d}\sum_{j}   \int_{\rm BZ} \!\!\!\! d \k \; f_{j\k}   \ev{\partial_{k_\alpha}   u_{j\k}  | \, U_{j\k}(t) \, |\partial_{k_\beta}  u_{j\k}} , \nn U_{j\k}(t)  &=& \theta(t) \sum_{\epsilon_{j'\k} > \epsilon_{\rm F}} \ket{ u_{j'\k}}  \emi{\omega_{jj'\k} t}\bra{ u_{j'\k}} . \label{02} \eea In the insulating case this is apparently similar to---and conceptually different from---the time-dependent quantum geometric tensor of Refs. \cite{Komissarov24,Verma25}.

By construction, at  time $t= 0^+$ \equ{02} yields the standard band-structure expression for the metric-curvature tensor  evaluated in the parallel-transport gauge \cite{Vanderbilt}:  \[ F_{\alpha\beta}(0^+) = \frac{\vc}{(2\pi)^d}\sum_{j}   \int_{\rm BZ} \!\!\!\! d \k \; f_{j\k}   \ev{\partial_{k_\alpha}   u_{j\k}   |\partial_{k_\beta}  u_{j\k}} ; \label{swm} \] this is subsequently evolved by the unitary matrix $ U_{j\k}(t) $. Notice that \equ{swm} applies even to the metallic case: its imaginary part is then proportional to the intrinsic anomalous Hall conductivity \cite{Nagaosa10}, while its real part is proportional to the Souza-Wilkens-Martin integral \cite{Souza00,rap_a33}, where only the regular part of the longitudinal conductivity is included.

A final comment concerns dissipation. Insofar as $L$ stays finite, $\sigma_{\alpha\beta}(\omega)$ is causal but nondissipative. It becomes dissipative after the $L \rightarrow \infty$ limit is performed, because an infinite system can be regarded as its own thermostat \cite{GiulianiVignale}. Owing to Ohm's law the Joule heating $Q$ is proportional to the squared electric field \cite{LandauEM}. When the spectrum is continuous---as in \equ{t3}---its geometrical  expression in terms of  ${\cal F}_{\alpha\beta}(\omega) $ becomes \cite{Nagaosa10}:
\bea Q &=&  \frac{e^2 \omega}{2\hbar  L^d}  \left\{ \mbox{Re }{\cal F}^{(+)}_{\alpha\beta}(\omega) \, \mbox{Re } [\EE_\alpha(\omega)  \EE_\beta^*(\omega)]  \right. \nn &+& \left. \mbox{Im }{\cal F}^{(-)}_{\alpha\beta}(\omega) \, \mbox{Im } [\EE_\alpha(\omega)  \EE_\beta^*(\omega)] \right\} . \eea

\section{Conclusions}

 I have shown that both the metric and the curvature are naturally endowed with a time dependence, which elucidates the geometrical nature of electron transport even beyond the adiabatic regime.
The causal evolution of the metric is at the root of the longitudinal optical conductivity and provides its explicit exact expression at any frequency, with the exception of the Drude term in the metallic case. Analogously, the evolution of the curvature is at the root of the transverse optical conductivity and provides its explicit exact expression.  The presentation is made in a many-body framework, 
{\rr which shows the main concepts by means of compact and transparent notations and formula\ae, with the additional virtue of formally dealing with interacting and disordered systems on the same ground.  As in the adiabatic case, the conversion of the  many-body quantum geometry into the more familiar Bloch geometry \cite{Vanderbilt,rap_a37} is straightforward by following a set of simple rules \cite{rap165}; the present nonadiabatic case is demonstrated in detail.}

I thank Raquel Queiroz for the fruitful conversations we had about this topic.
Work supported by the Office of Naval Research (USA) Grant No. N00014-20-1-2847.

\appendix
\section{Kubo formul\ae }

I adopt here the notations of the main text in order to explicitly display the relevant Kubo formul\ae.

\smallskip\noindent
{\it Longitudinal (symmetric) conductivity}: \[  \sigma_{\alpha\beta}^{(+)}(\omega) = D_{\alpha\beta} \left[ \delta(\omega) + \frac{i}{\pi \omega} \right] +\sigma_{\alpha\beta}^{(\rm regular)}(\omega) , \label{cond} \] 
\index{conductivity}
\[ D_{\alpha\beta} = \frac{\pi e^2}{ L^d} \left( \frac{N}{m} \delta_{\alpha\beta} - \frac{2}{\hbar} {\sum_{n\neq 0}}  \frac{{\cal R}_{n,\alpha\beta}  }{\omega_{0n}} \right) \label{drude} , \]
where for $ \omega>0$:
\bea \mbox{Re } \sigma_{\alpha\beta}^{(\rm regular)}(\omega) &=& \frac{\pi e^2}{\hbar L^d}  {\sum_{n\neq 0}} \frac{ {\cal R}_{n,\alpha\beta}}{\omega_{0n}} \delta(\omega - \omega_{0n}) , \; \label{s1} \\ \mbox{Im } \sigma_{\alpha\beta}^{(\rm regular)}(\omega) &=& \frac{2 e^2}{\hbar L^d}  {\sum_{n\neq 0}} \frac{ {\cal R}_{n,\alpha\beta}}{\omega_{0n}} \frac{\omega}{\omega_{0n}^2 - \omega^2} \label{s2} ; \eea  the Drude weight $D_{\alpha\beta}$ vanishes in insulators. {\rr \equ{drude} can be actually summed; the result coincides with \equ{omega1} above, where  it is alternatively derived as the nongeometrical term in the adiabatic response.}

The real part of longitudinal conductivity obeys the $f$-sum rule  \bea \int_0^\infty d \omega \; \mbox{Re } \sigma_{\alpha\beta} (\omega) &=& \frac{D_{\alpha\beta}}{2} + \int_0^\infty d \omega \; \mbox{Re } \sigma_{\alpha\beta}^{(\rm regular)} (\omega) \nn &=& \frac{\pi e^2 N}{2 m L^d}\delta_{\alpha\beta} , \label{fsum} \eea

\smallskip\noindent
{\it Transverse (antisymmetric) conductivity:} \\
The intrinsic (geometric) contribution to the anomalous Hall conductivity \cite{Nagaosa10} requires spontaneous breaking of time-reversal symmetry. Its expression is:
\bea \mbox{Re } \sigma_{\alpha\beta}^{(-)}(\omega) &=& \frac{2e^2}{\hbar  L^d} {\sum_{n\neq 0}} \frac{{\cal I}_{n,\alpha\beta}}{\omega_{0n}^2 - \omega^2} \label{s3}
\\ \mbox{Im } \sigma_{\alpha\beta}^{(-)}(\omega) &=& \frac{\pi e^2}{\hbar L^d} {\sum_{n
\neq 0}} \frac{{\cal I}_{n,\alpha\beta}}{\omega_{0n}} \delta(\omega - \omega_{0n}) , \; \omega> 0 . \label{s4} \eea 
The dc transverse conductivity is easily recast in terms of the many-body Berry curvature at $\kk=0$: \[ \mbox{Re } \sigma_{\alpha\beta}^{(-)}(0) = - \frac{e^2}{\hbar  L^d}  \Omega_{\alpha\beta} .  \label{cc} \] The expression holds for metals and insulators, in either 2$d$ or 3$d$.

\smallskip\noindent
{\it Polarizability (symmetric):} \\
The polarizability tensor in insulators is defined as $\chi_{\alpha\beta}(\omega) = \partial P_\alpha(\omega)/\partial \EE_\beta(\omega)$. Since the current is the time derivative of the polarization, one has $ j_\alpha(\omega) = -i\omega P_\alpha(\omega) = -i \omega \chi_{\alpha\beta}(\omega) \EE_\beta(\omega) $ and \[ \mbox{Re } \chi_{\alpha\beta}(\omega) = \frac{\mbox{Im }\sigma^{(+)}_{\alpha\beta}(\omega)}{\omega} = \frac{2 e^2}{\hbar L^d}  {\sum_{n\neq 0}} \frac{ {\cal R}_{n,\alpha\beta}}{\omega_{0n}(\omega_{0n}^2 - \omega^2)} .\] In the $\omega \rightarrow 0$ limit one retrieves the Kubo formula for the static polarizability: \[  \chi_{\alpha\beta}(0) =   \frac{2  e^2}{\hbar L^d}  {\sum_{n\neq 0}} \frac{  \ev{\Psi_0| \hat{v}_\alpha | \Psi_n} \ev{\Psi_n| \hat{v}_\beta |\Psi_0} }{\omega_{0n}^3} , \label{ss} \]


\begin{thebibliography}{10}

\bibitem{Berry84}
{ M. V. Berry, Quantal phase factors accompanying adiabatic changes, Proc. Roy. Soc. Lond. A {\bf 392}, 45 (1984)}.

\bibitem{Niu84}
{ Q. Niu and D. J. Thouless, Quantised adiabatic charge transport in the presence of
substrate disorder and many-body interaction, J. Phys A {\bf 17}, 2453 (1984)}.

\bibitem{Vanderbilt}
{ D. Vanderbilt, {\it Berry Phases in Electronic Structure Theory} (Cambridge
  University Press, Cambridge, 2018)}.

\bibitem{Xiao10}
{ D. Xiao, M.-C. Chang, and Q. Niu, Berry phase effects on electronic properties, Rev. Mod. Phys. {\bf 82}, 1959 (2010)}.

\bibitem{Nagaosa10}
{ N. Nagaosa, J. Sinova, S. Onoda, A. H. MacDonald, and N. P. Ong, Anomalous Hall effect, Rev. Mod.
  Phys. {\bf 82}, 1539 (2010)}.

\bibitem{rap_a37}
{ R. Resta, Berry phase and geometrical observables, in: {\it Encyclopedia of Condensed Matter Physics, 2nd edition}, T.
  Chakraborty ed.\ (Elsevier, 2023), p. 670}.

\bibitem{rap168}
{ R. Resta, Molecular Berry curvatures and the adiabatic response tensors, J. Chem. Phys. {\bf 158}, 024105 (2023)}.

\bibitem{Ahn22}
{ J. Ahn, G.-Y. Guo, N. Nagaosa, and A. Vishwanath, Riemannian geometry of resonant optical responses, Nat. Phys. {\bf 18}, 290
  (2022)}.

\bibitem{Nagaosa22}
{ N. Nagaosa, Nonlinear optical responses in noncentrosymmetric quantum materials, Annals of Physics {\bf 447}, 169146 (2022)}.

\bibitem{Komissarov24}
{ I. Komissarov, T. Holder, and R. Queiroz, The quantum geometric origin of capacitance in insulators, Nat. Commun. {\bf 15}, 4621 (2024)}.

\bibitem{Verma25}
{ N. Verma and R. Queiroz, Instantaneous response and quantum geometry of insulators, {\tt https://arxiv.org/pdf/2403.07052}}.

\bibitem{Ortiz94}
{ G. Ort\'{\i}z and R. M. Martin, Macroscopic polarization as a geometric quantum phase: Many-body formulation, Phys. Rev. B {\bf 49}, 14202 (1994)}.

\bibitem{rap165}
{ R. Resta, Theory of longitudinal and transverse nonlinear dc conductivity, Phys. Rev. Research {\bf 4}, 033002 (2022)}.

\bibitem{rap169}
{ R. Resta, Geometrical theory of the shift current in presence of disorder and interaction, Phys. Rev. Lett. {\bf 133}, 206903 (2024)}.

\bibitem{rap135}
{ D. Ceresoli and R. Resta, Orbital magnetization and Chern number in a supercell framework: Single k-point formula, Phys. Rev. B {\bf 76}, 012405 (2007)}.

\bibitem{Souza00}
{ I. Souza, T. Wilkens, and R. M. Martin, Polarization and localization in insulators: Generating function approach, Phys. Rev. B {\bf 62}, 1666 (2000)}.

\bibitem{rap_a33}
{ R. Resta, Theory of the insulating state, Riv. Nuovo Cimento {\bf 41}, 463 (2018)}.

\bibitem{Kato50}
{ T. Kato, On the adiabatic theorem of quantum mechanics, J. Phys. Soc. Jpn. {\bf 5}, 435 (1950)}.

\bibitem{Kohn64}
{ W. Kohn, Phys. Theory of the insulating state, Rev. {\bf 133}, {A171} (1964)}.

\bibitem{GiulianiVignale}
{ G. F. Giuliani and G. Vignale, {\it Quantum Theory of the Electron Liquid}
  (Cambridge University Press, Cambridge, 2005)}.

\bibitem{LandauEM}
{ L. D. Landau, E. N. Lifshitz, and L. Pitaevskii, {\it Electrodynamics of
  Continuous Media}, 2nd Ed. (Elsevier, Oxford, 2008)}.

\end{thebibliography}

\end{document}